\documentclass{article}
\begin{document}
\title{Coherent States and Duality}
\author{Jos\'e M. Isidro \\ 
Department of Theoretical Physics,\\ 
1 Keble Road, Oxford OX1 3NP, UK\\ 
and\\
Instituto de F\'{\i}sica Corpuscular (CSIC-UVEG)\\
Apartado de Correos 22085, 46071 Valencia, Spain\\
{\tt isidro@thphys.ox.ac.uk}}
\maketitle

\begin{abstract}

We formulate a relation between quantum--mechanical coherent states 
and complex--differentiable structures on the classical phase space ${\cal C}$
of a finite number of degrees of freedom. Locally--defined coherent 
states parametrised by the points of ${\cal C}$ exist when there is
an almost complex structure on ${\cal C}$. When ${\cal C}$ admits a complex 
structure, such coherent states are globally defined on ${\cal C}$.
The picture of quantum mechanics that emerges allows to implement duality 
transformations.

2001 Pacs codes: 03.65.Bz, 03.65.Ca, 03.65.-w.

\end{abstract}

\section{Introduction}\label{intro}

Duality in strings, branes and M--theory \cite{DUALITY, LAG} motivates the study of duality 
in quantum mechanics. Now scattering amplitudes in perturbative string theory 
can be organised in a manner analogous to the loop expansion of quantum field theory.
The latter expansion is a series in powers of $\hbar$, with an $L$--loop 
amplitude contributing a prefactor $\hbar^{L-1}$. In string theory, the 
role of loops is played by the genus of the string worldsheet, and the 
role of $\hbar$ is played by the string coupling $g_s$. By the same token, 
duality in quantum mechanics refers to different regimes in a series 
expansion in powers of $\hbar$.

In the standard formulation of quantum mechanics, 
if one observer calls a certain phenomenon {\it semiclassical}, 
then so will it be for all other observers. If one observer calls 
a certain phenomenon {\it strong quantum}, then so will it be for all other observers.
This does not allow for the relativity of the concept of a quantum
that underlies the notion of duality. In view of these developments, 
a framework for quantum mechanics is required that can accommodate such a relativity 
of of the concept of a quantum. In this letter we address this issue through an analysis 
of coherent states \cite{COHST}.

\section{Global coherent states}\label{globcoh}

Throughout this letter, ${\cal C}$ will denote a $2n$--dimensional classical 
phase space endowed with a symplectic form $\omega_{\cal C}$ that, in local Darboux 
coordinates, may be written as
\begin{equation}
\omega_{\cal C}=\sum_{l=1}^n{\rm d} p_l \wedge {\rm d} q^l.
\label{cano}
\end{equation}
Let us assume that ${\cal C}$ admits a complex structure 
${\cal J}_{\cal C}$. Furthermore let ${\cal J}_{\cal C}$ be compatible with the symplectic 
structure $\omega_{\cal C}$. This means that the real and imaginary parts of the holomorphic 
coordinates $z^l$ for ${\cal J}_{\cal C}$ are Darboux coordinates for the symplectic form 
$\omega_{\cal C}$:
\begin{equation}
z^l=q^l + {\rm i} p_l,  \qquad l=1,\ldots, n.
\label{compcoords}
\end{equation}
The set of all $z^l$ so defined provides a holomorphic atlas for ${\cal C}$.
Upon quantisation, the Darboux coordinates
$q^l$ and $p_l$ become operators $Q^l$ and $P_l$ satisfying the Heisenberg algebra 
\begin{equation}
[Q^j, P_k]={\rm i}\delta^j_k.
\label{hei}
\end{equation}
Define the annihilation operators
\begin{equation}
A^l=Q^l+{\rm i} P_l,\qquad l=1,\ldots, n.
\label{anni}
\end{equation}
Quantum excitations are measured with respect to a vacuum state 
$|0\rangle$. The latter is defined as that state in the Hilbert space 
${\cal H}$ which satisfies
\begin{equation}
A^l|0\rangle = 0,  \qquad l=1, \ldots, n,
\label{vac}
\end{equation}
and coherent states $|z^l\rangle$ are eigenvectors of $A^l$, 
with eigenvalues given in equation (\ref{compcoords}) above:
\begin{equation}
A^l|z^l\rangle=z^l|z^l\rangle,\qquad l=1,\ldots, n.
\label{annop}
\end{equation}

How do the vacuum state $|0\rangle$ and the coherent states $|z^l\rangle$
transform under a canonical coordinate transformation on ${\cal C}$? 
Call the new Darboux coordinates $q'^l$, $p'_l$.  Upon quantisation the 
corresponding operators $Q'^l$, $P'_l$ continue to satisfy the Heisenberg 
algebra (\ref{hei}). Then the combinations
\begin{equation}
z'^l=q'^l + {\rm i} p'_l,  \qquad l=1,\ldots, n
\label{xcompcoords}
\end{equation}
continue to provide holomorphic coordinates for ${\cal C}$, and the 
transformation between the $z^l$ and the $z'^l$ is given by an $n$--variable
holomorphic function $f$,
\begin{equation}
z'=f(z),\qquad \bar\partial f=0.
\label{fhol}
\end{equation}
We can write as above
\begin{equation}
A'^l=Q'^l+{\rm i} P'_l,\qquad l=1,\ldots, n,
\label{annix}
\end{equation}
\begin{equation}
A'^l|0\rangle = 0,  \qquad l=1, \ldots, n,
\label{vacx}
\end{equation}
\begin{equation}
A'^l|z'^l\rangle=z'^l|z'^l\rangle,\qquad l=1,\ldots, n.
\label{annopx}
\end{equation}
There is no physical difference between equations (\ref{anni}), (\ref{vac}) and 
(\ref{annop}), on the one hand, and their holomorphic transforms (\ref{annix}), 
(\ref{vacx}) and (\ref{annopx}), on the other. Under the transformation 
(\ref{fhol}), the vacuum state $|0\rangle$ is mapped into itself, 
and the coherent states $|z^l\rangle$ are mapped into the coherent states $|z'^l\rangle$.
Therefore the notion of coherence is global for all observers on ${\cal 
C}$, {\it i.e.}, any two observers will agree on what is a coherent 
state {\it vs.} what is a noncoherent state. A consequence of 
this fact is the following. Under holomorphic diffeomorphisms of ${\cal C}$, 
the semiclassical regime of the quantum theory on ${\cal H}$ is mapped 
into the semiclassical regime, and the strong quantum regime is mapped 
into the strong quantum regime.

\section{Local coherent states}\label{loccoh}

We now relax the conditions imposed on ${\cal C}$. In this section we will assume
that ${\cal C}$ carries an almost complex structure $J_{\cal C}$ compatible with 
the symplectic structure $\omega_{\cal C}$. Specificallly, an almost complex structure 
is defined as a tensor field $J_{\cal C}$ of type $(1,1)$ such that, at every point of 
${\cal C}$, $J_{\cal C}^2=-{\bf 1}$.  Using Darboux coordinates $q^l$, 
$p_l$ on ${\cal C}$ let us form the combinations
\begin{equation}
w^l=q^l + {\rm i} p_l, \qquad l=1,\ldots, n.
\label{combw}
\end{equation} 
Compatibility between $\omega_{\cal C}$ and $J_{\cal C}$ means that we can take $J_{\cal C}$ 
to be 
\begin{equation}
J_{\cal C}\left({\partial\over\partial w^l}\right)={\rm i}{\partial\over\partial w^l},
\qquad 
J_{\cal C}\left({\partial\over\partial \bar w^l}\right)=-{\rm i}{\partial\over\partial\bar w^l}.
\label{jota}
\end{equation}
Unless ${\cal C}$ is a complex  manifold to begin with, equations 
(\ref{combw}) and (\ref{jota}) fall short of defining a complex structure ${\cal J}_{\cal C}$. 
The set of all such $w^l$ does not provide a holomorphic atlas for ${\cal C}$. 
There exists at least one canonical coordinate transformation between 
Darboux coordinates, call them $(q^l, p_l)$ and $(q'^l, p'_l)$, such that 
the passage between $w^l=q^l+{\rm i}p_l$ and $w'^l=q'^l+{\rm i}p'_l$
is given by a nonholomorphic function $g$ in $n$ variables,
\begin{equation}
w'=g(w,\bar w), \qquad \bar\partial g\neq 0.
\label{nonhol}
\end{equation}

Mathematically, nonholomorphicity implies the mixing of $w^l$ and $\bar w^l$.
Quan-tum--mechanically, the loss of holomorphicity has deep physical consequences.
One would write, in the initial coordinates $w^l$, a defining equation for the vacuum 
state $|0\rangle$
\begin{equation}
a^l|0\rangle =0, \qquad l=1,\ldots, n,
\label{xvac}
\end{equation}
where $a^l=Q^l + {\rm i} P_l$ is the corresponding local annihilation operator.
However, one is just as well entitled to use the new coordinates $w'^l$
and write
\begin{equation}
a'^l|0'\rangle =0, \qquad l=1,\ldots, n,
\label{xvacu}
\end{equation}
where we have primed the new vacuum, $|0'\rangle$. 
Are we allowed to identify the states $|0\rangle$ and $|0'\rangle$? 
We could identify them if $w'^l$ were a holomorphic function of $w^l$; 
such was the case in section \ref{globcoh}. However, now we are considering 
a nonholomorphic transformation, and we cannot remove the 
prime from the state $|0'\rangle$. This is readily seen.
We have
\begin{equation}
a'=G(a, a^{\dagger}),
\label{effe}
\end{equation}
with $G$ a quantum nonholomorphic function corresponding to the classical 
nonholomorphic function $g$ of equation (\ref{nonhol}). As $[a^j,a_k^{\dagger}]=\delta^j_k$, 
ordering ambiguities will arise in the construction of $G$ from $g$,
that are usually dealt with by normal ordering. Normal ordering would 
appear to allow us to identify the states $|0\rangle$ and $|0'\rangle$.
However this is not the case, as there are choices of $g$ that are left invariant under 
normal ordering, such as the sum of a holomorphic function plus
an antiholomorphic function, $g(w,\bar w) = g_1(w) + g_2(\bar w)$.
Under such a transformation one can see that the state $|0\rangle$ satisfying 
eqn. (\ref{xvac}) will not satisfy eqn. (\ref{xvacu}). 
We conclude that, in the absence of a complex structure 
on classical phase space, the vacuum depends on the observer. 
The state $|0\rangle$ is only defined locally on ${\cal C}$; it cannot be 
extended globally to all of ${\cal C}$. 

Similar conclusions may be expected for the coherent states $|w^l\rangle$. 
The latter are defined only locally, as eigenvectors of the local annihilation 
operator, with eigenvalues given in equation (\ref{combw}):
\begin{equation}
a^l|w^l\rangle=w^l|w^l\rangle,\qquad l=1,\ldots, n.
\label{annopp}
\end{equation}
Due to $[a^j,a_k^{\dagger}]=\delta^j_k$, under the nonholomorphic 
coordinate transformation (\ref{nonhol}), the local coherent states 
$|w^l\rangle$ are {\it not} mapped into the local coherent states 
satisfying
\begin{equation}
a'^l|w'^l\rangle=w'^l|w'^l\rangle,\qquad l=1,\ldots, n
\label{annoppp}
\end{equation}
in the primed coordinates. No such problems arose for the holomorphic operator 
equation $A'=F(A)$ corresponding to the holomorphic coordinate change $z'=f(z)$ 
of equation (\ref{fhol}), because the commutator $[A^j, A_k^{\dagger}]=\delta^j_k$ 
played no role. Thus coherence becomes a local property on classical phase space.
In particular,  observers on ${\cal C}$ not connected by means of a holomorphic change 
of coordinates need not, and in general will not, agree on what is a semiclassical 
effect {\it vs.} what is a strong quantum effect. 

\section{Coherence}\label{prov}

We have called {\it coherent} the states constructed in previous sections.
However, we have not verified that they actually satisfy the usual requirements 
imposed on coherent states \cite{COHST}. What ensures that the states so 
constructed are actually coherent is the following argument. 

We have made no reference to coupling constants or potentials, with the understanding 
that the Hamilton--Jacobi method has already placed us, by means of suitable coordinate 
transformations, in a coordinate system on ${\cal C}$ where all interactions 
vanish \cite{ARNOLD}.
Then any dynamical system with $n$ independent degrees of freedom that can be transformed 
into the freely--evolving system can be further mapped into the $n$--dimensional harmonic 
oscillator. The combined transformation is canonical. Moreover it is locally holomorphic when 
${\cal C}$ is an almost complex manifold. Thus locally on ${\cal C}$, our global states 
$|z^l\rangle$ of section \ref{globcoh} coincide with the coherent states of the 
$n$--dimensional harmonic oscillator. Mathematically this fact reflects the structure 
of a complex manifold: locally it is always holomorphically diffeomorphic to 
${\bf C}^n$ \cite{ARNOLD}. Physically this fact reflects 
the decomposition into the creation and annihilation modes of perturbative quantum 
mechanics and field theory. In this way, the mathematical problem of patching together different 
local coordinates $ z^l_{\alpha}$ labelled by an index 
$\alpha$ may be recast in physical terms. It is the patching together of different 
local expansions into creators $A^{\dagger}_{\alpha}$ and annihilators $A_{\alpha}$, 
for different values of $\alpha$.

In particular, we can write the resolution of unity on ${\cal H}$ associated 
with a holomorphic atlas on ${\cal C}$ consisting of coordinates $z^l_{\alpha}$:
\begin{equation}
\sum_{\alpha}\sum_{l=1}^n\int_{\cal C} {\rm d}\mu_{\cal C}\, |z^l_{\alpha}\rangle  \langle 
z^l_{\alpha}|= {\bf 1},
\label{res}
\end{equation}
where ${\rm d}\mu_{\cal C}$ is an appropriate measure on ${\cal C}$. 

Analogous arguments are also applicable to the local states $|w^l\rangle$ 
of section \ref{loccoh}. 
In particular, every almost complex manifold is locally a complex manifold.
Every holomorphic coordinate chart on ${\cal C}$ is  diffeormorphic to 
${\bf C}^n$, so the $|w^l\rangle$ look locally like the coherent 
states of the $n$--dimensional harmonic oscillator.
However, the loss of holomorphicity of ${\cal C}$ 
alters equation (\ref{res}) in one important way. We may write as above
\begin{equation}
\sum_{\alpha}\sum_{l=1}^n\int_{\cal C} {\rm d}\mu_{\cal C}\, |w^l_{\alpha}\rangle  \langle 
w^l_{\alpha}|,
\label{resx}
\end{equation}
but we can no longer equate this to the identity on 
${\cal H}$. The latter is a {\it complex}\/ vector space, while eqn.
(\ref{resx}) allows one at most to expand an arbitrary,
real--analytic function on ${\cal C}$, since the latter is just a 
real--analytic manifold. Hence we cannot equate (\ref{resx})
to ${\bf 1}_{\cal H}$. We can only equate it to the identity on the {\it 
real}\/
Hilbert space of real--analytic functions on ${\cal C}$.
This situation is not new; coherent states without a resolution of unity have been 
analysed in ref. \cite{NOUNITY}, where they have been related to the choice of 
an inadmissible fiducial vector.  

\section{Discussion}\label{discussion}

The purpose of this article is to present a framework for quantum mechanics in which
coherent states are defined locally on classical phase space ${\cal C}$, 
but not necessarily globally. This is an explicit implementation of the relativity 
(underlying the notion of duality) of the concept of a quantum.

Our results may be summarised as follows. Coherent states can always 
be defined locally, {\it i.e.}, in the neighbourhood of any point  on ${\cal C}$. 
When there is a complex structure ${\cal J}_{\cal C}$, coherence becomes 
a global property on ${\cal C}$. 
In the absence of a complex structure, however, the best we can do 
is to combine Darboux coordinates $q$, $p$ as $q+{\rm i}p$. 
Technically this only defines an almost complex structure $J_{\cal C}$ on ${\cal C}$.
Since the combination $q+{\rm i}p$ 
falls short of defining a complex structure, quantities depending on $q+{\rm i}p$ 
on a certain coordinate patch will generally also depend on $q-{\rm i}p$ 
when transformed to another coordinate patch. This proves that coherence remains 
a local property on classical phase space: observers not connected by means of 
a holomorphic change of coordinates need not, and in general will not,
agree on what is a semiclassical effect {\it vs.} what is a strong quantum effect. 

Most physical systems admit a complex structure ${\cal J}_{\cal C}$ on their 
classical phase spaces ${\cal C}$. Prominent among them is the 
1--dimensional harmonic oscillator. Mathematically, the corresponding 
${\cal C}$ supports the simplest holomorphic structure, that of the complex plane.
Physically, canonical quantisation rests on the decomposition 
of a field into an infinite number of oscillators. The notions that the vacuum state 
is unique, and that coherence is a universal property independent of the 
observer, follow naturally. However,  
recent breakthroughs in quantum field theory and M--theory suggest the need 
for a framework in which duality transformations can be accommodated at the elementary 
level of quantum mechanics, before considering field theory or strings. 
This, in turn, would help to understand better the dualities underlying 
quantum fields, strings and branes.

The formalism presented here can accommodate duality transformations in a 
natural way. In the absence of a complex structure ${\cal J}_{\cal C}$, 
all our statements concerning the vacuum state and the property of coherence are 
necessarily local in nature, {\it i.e.}, they do not hold globally on ${\cal C}$. 
A duality transformation of the quantum theory on ${\cal H}$ will thus be specified 
by a nonholomorphic coordinate transformation on ${\cal C}$.

However, the question immediately arises: do we not have an overabundance 
of vacua? Does every imaginable nonholomorphic transformation induce a 
{\it physical}\/ duality? 
Duality is not to be understood as a transformation between different physical 
phenomena. Rather, it is to be understood as a transformation between different 
descriptions of the same quantum physics.
A judicious application of physical symmetries 
can vastly restrict this apparent overabundance of vacua. 
Usually dualities appear under the form 
of a group ${\cal D}$. Rather than taking every imaginable nonholomorphic transformation 
to define a physical duality we must assume, as is the case in M--theory, 
a knowledge of the duality group ${\cal D}$,  
and restrict ourselves to those nonholomorphic transformations that actually realise it. 

In using coherent states our analysis has been geometric. Geometric 
approaches to quantum mechanics \cite{JACKIW, ANANDAN, GQCS, MATONE, MEX} 
have proved extremely useful. In particular, 
that coherence equals holomorphicity has been known for long \cite{LONG}. 
Here we have proved that noncomplex structures (such as almost complex structures) 
allow to implement duality transformations.

{\bf Acknowledgements}

Support from PPARC (grant PPA/G/O/2000/00469) and DGICYT (grant PB 96-0756) is acknowledged.


\begin{thebibliography}{99}

\bibitem{DUALITY}
M. Kaku, {\it Strings, Conformal Fields and M--Theory},  Springer, Berlin 
(2000).

\bibitem{LAG}
L. Alvarez--Gaum\'e and S. Hassan, {\it Fortsch. Phys.} {\bf 45} (1997) 159.

\bibitem{COHST}   
J. Klauder and B.--S. Skagerstam, {\it Coherent States}, World Scientific, Singapore 
(1985). For recent reviews see J. Klauder, {\tt quant-ph/9810043}, {\tt quant-ph/0110108}.

\bibitem{ARNOLD}
V. Arnold, {\it Mathematical Methods of Classical Mechanics}, Springer, Berlin (1989).

\bibitem{NOUNITY}
J. Klauder, {\tt quant-ph/0008132}.

\bibitem{JACKIW}
L. Faddeev and R. Jackiw, {\it Phys. Rev. Lett.} {\bf 60} (1988) 1692;
R. Jackiw, {\it Constrained Quantization without Tears}, in {\it Constraint 
Theory and Quantization Methods}, F. Colmo {\it et al.} (eds.), World 
Scientific, Singapore (1994); R. Jackiw, {\it Diverse Topics in 
Theoretical and Mathematical Physics}, World Scientific, Singapore (1995).

\bibitem{ANANDAN}
J. Anandan and Y. Aharonov, {\it Phys. Rev. Lett.} {\bf 65} (1990) 1697;
J. Anandan, {\it Found. Phys.} {\bf 21} (1991) 1265; {\tt quant-ph/0012011}.

\bibitem{GQCS}
J. Klauder, {\tt gr-qc/0112053},  {\tt quant-ph/0112010}.

\bibitem{MATONE}  
A. Faraggi and M. Matone,
{\it Int. J. Mod. Phys.} {\bf A15} (2000) 1869;
G. Bertoldi, A. Faraggi and M. Matone, 
{\it Class. Quant. Grav.} {\bf 17} (2000) 3965.

\bibitem{MEX}
H. Garc\'{\i}a--Compe\'an, J. Pleba\'nski, M. Przanowski and F. 
Turrubiates,  {\it Int. J. Mod. Phys.} {\bf A16} (2001) 2533;
{\tt hep-th/0112049}; I. Carrillo--Ibarra and H. Garc\'{\i}a--Compe\'an, 
{\tt hep-th/0202015}.

\bibitem{LONG}
J. Klauder, {\it Ann. Phys.} {\bf 11} (1960) 123;
V. Bargman, {\it Comm. Pure. Appl. Math.} {\bf 14} (1961) 187.

\end{thebibliography}
\end{document}